\documentclass[apjl,twocolumn]{openjournal}
\usepackage{natbib} 
\usepackage[dvipsnames]{xcolor}
\usepackage{aas_macros} 
\usepackage{amssymb}
\usepackage{amsmath}
\usepackage{threeparttable}
\usepackage[title]{appendix}
\usepackage{hyperref}	
\hypersetup{colorlinks=true,linkcolor=blue,citecolor=blue,filecolor=blue,urlcolor=blue}
\usepackage[caption=false]{subfig}
\usepackage{soul}                          

\begin{document}
\title[]{Early Bright Galaxies from Helium Enhancements in High-Redshift Star Clusters\vspace{-15mm}}
\author{Harley Katz$^{1,2}$\thanks{$^*$E-mail: \href{mailto:harleykatz@uchicago.edu}{harley.katz@physics.ox.ac.uk}}, Alexander P. Ji$^{1,2}$, O. Grace Telford$^{3,4}$\thanks{Carnegie-Princeton Fellow}, \& Peter Senchyna$^{4}$\thanks{NHFP Hubble Fellow}
}

\affiliation{$^{1}$Department of Astronomy \& Astrophysics, University of Chicago, 5640 S Ellis Avenue, Chicago, IL 60637, USA}
\affiliation{$^2$Kavli Institute for Cosmological Physics, University of Chicago, Chicago, IL 60637, USA}
\affiliation{$^{3}$Department of Astrophysical Sciences, Princeton University, 4 Ivy Lane, Princeton, NJ 08544, USA}
\affiliation{$^{4}$The Observatories of the Carnegie Institution for Science, 813 Santa Barbara Street, Pasadena, CA 91101, USA}

\begin{abstract}
The first few cycles of JWST have identified an overabundance of UV-bright galaxies and a general excess of UV luminosity density at $z\gtrsim10$ compared to expectations from most (pre-JWST) theoretical models. Moreover, some of the brightest high-redshift spectroscopically confirmed galaxies exhibit peculiar chemical abundance patterns, most notably extremely high N/O ratios. Since N/O has been empirically shown to scale strongly with He/H, as expected for hot hydrogen burning, these same bright high-redshift galaxies are likely also helium-enhanced. Under simplistic assumptions for stellar evolution, the bolometric luminosity of a star scales as $L\propto(2-\frac{5}{4}Y)^{-4}(2-Y)^{-1}$ --- hence a higher He/H leads to brighter stars. In this Letter, we evolve a series of {\small MESA} models to the zero-age main-sequence and highlight that the helium enhancements at the levels measured and inferred for high-redshift galaxies can boost the 1500~\AA\ UV luminosity by up to $\sim50\%$, while simultaneously increasing the stellar effective temperature. The combination of helium enhancements  with nebular continuum emission expected for intense bursts of star formation have the potential to help reduce the tension between JWST observations and certain galaxy formation models. 
\end{abstract}

\section{Introduction}
\label{sec:intro}
Photometric and spectroscopic measurements from JWST \citep{Garnder2006} have demonstrated that there are both more UV-bright galaxies at high redshift and more generally an excess UV luminosity density compared to expectations from most pre-JWST theoretical models of high-redshift galaxy formation \citep[e.g.,][although cf. \citealt{Willott2023}]{Finkelstein2023_b,Harikane2024,Leung2023,Chemerynska2023}. Numerous astrophysical solutions have been proposed to explain these observations that can be categorized as either (1) modifications to the observed mass-to-light ratio or (2) appealing to a higher star formation efficiency.

Within the first category, possible physical mechanisms that can impact the mass-to-light-ratio include (a) assuming the scatter between halo mass and UV magnitude increases at high redshift, possibly due to large fluctuations in the star formation rate \citep[e.g.,][]{Ren2019,Mason2023,Shen2023,Sun2023,Kravtsov2024}, (b) a decrease in the dust attenuation of high-redshift massive galaxies as a result of strong radiation pressure \citep{Ferrara2023}, (c) modifying the stellar initial mass function by, for example, making it more top-heavy (e.g. \citealt{Yung2024}, although cf. \citealt{Cueto2024}), and (d) an increased role of nebular continuum emission due to, for example, intense star formation events and/or hotter or more massive stars \citep{Katz2024_BJ}. Alternatively, some simulations and models demonstrate that the star formation efficiency may be higher in the early Universe due to weaker feedback and strong gas inflows \citep[e.g.,][]{McCaffrey2023,Dekel2023}.

The variety of proposed solutions are not mutually exclusive --- the underlying physical origin of both the excess UV luminosity density and number density of high-redshift bright galaxies is likely a combination of multiple factors. However, theoretical predictions for UV luminosities inevitably rely on stellar population synthesis (SPS) models such as {\small STARBURST99} \citep{Leitherer1999}, {\small BPASS} \citep{Stanway2018}, or {\small BC03} \citep{Bruzual2003}. The predicted luminosities of the SPS models are subject to certain underlying assumptions, for example, chemical abundance ratios. It is typically assumed that metal abundance ratios are independent of metallicity and that the helium abundance is close to the value of $Y=0.241$ as measured from the CMB \citep{Planck2020} with perhaps a slight excess that increases with metallicity. 

However, spectroscopic observations from JWST have revealed that the chemical abundance patterns in numerous high-redshift galaxies are distinctly non-solar. Most-notably, a class of galaxies has been identified that exhibit strong rest-frame UV nitrogen emission lines \citep[e.g.,][]{Cameron2023_nit,Bunker2023,Isobe2023_nit,Senchyna2024,Castellano2024,Topping2024,Topping2024-NC,Schaerer2024}. Interestingly, three of these galaxies are at $z>9$, with GNz11 and GHZ2 being two of the brightest spectroscopically confirmed objects in the sky at these redshifts. Numerous studies have highlighted the similarity between the gas-phase nitrogen abundances in these high-redshift nitrogen-emitters and stellar abundance patterns found in local globular clusters \citep[e.g.,][]{Senchyna2024,Charbonnel2023,Marques-Chaves2024}. Assuming this connection is causal and based on stars in the Milky Way, \cite{Belokurov2023} estimated that $50-70\%$ of star formation in our galaxy likely occurs in massive, bound clusters with properties similar to globular clusters, prior to disk formation. 

One of the unique signatures of globular clusters is their peculiar chemical abundance patterns \citep[e.g.,][]{Bastian2018}. The fraction of nitrogen-enriched stars scales with the cluster initial mass \citep{Milone2017,Gratton2019,Belokurov2023,Usman2024} as does the enhancement in helium above the value expected from Big Bang nucleosynthesis \citep{Milone2018,Gratton2019}. For certain Milky Way globular clusters (e.g., Omega Centauri, NGC~2808) the helium mass fraction can reach as high as $Y\sim0.4$ \citep{Norris2004,Piotto2005,Dupree2013,Joo2013,Pasquini2011}. Other Milky Way globular clusters also show helium enhancements \citep[e.g.][]{Caloi2007,Busso2007,Dalessandro2013}, albeit less extreme than Omega Centauri and NGC~2808. Similar helium enhancements have also been observed in the Galactic bulge \citep{Renzini1994,Nataf2011}. Such helium enhancements up to $Y\sim0.4$ have now been observed at high redshift ($z\sim6$) in the gas phase alongside the aforementioned nitrogen enhancements \citep{Yanagisawa2024_He}, further tightening the link between these high-redshift nitrogen emitters and local globular clusters.

In this Letter, we simply highlight the fact that an increased helium mass fraction naturally increases both the luminosity and temperature of stars. If a large fraction of high-redshift star formation occurs in dense, bound clusters as suggested by both Milky Way data \citep{Belokurov2023} and high-redshift gravitationally lensed objects \citep[e.g.,][]{Vanzella2022,Adamo2024,Fujimoto2024,Mowla2024}, then significant helium enrichment and consequent luminosity enhancement may be a natural byproduct. Here, we quantify the excess luminosity one may expect for a helium-enriched population of stars and discuss how this may help relieve some of the tension regarding the excess UV luminosity at high redshift.


\section{The Impact of Helium Enhancements on 1500~\AA\ UV Luminosities}
\label{sec:intro}
It is well established that both the total bolometric luminosity of a star and its effective temperature are sensitive to both mean molecular weight ($\mu$) and opacity ($\kappa$). For a star in radiative equilibrium with an ideal gas equation of state, the bolometric luminosity ($L$) scales as (see, e.g., \citealt{Lamers2017}):
\begin{equation}
    L\propto\frac{\mu^4M^3}{\kappa}.
\end{equation}
For stars with $M>2\ {\rm M_{\odot}}$ where electron scattering represents the dominant opacity, $\kappa\propto1+X$, where $X$ is the mass fraction of hydrogen. Ignoring metals and assuming a fully ionized medium, $\kappa\rightarrow2-Y$ and $\mu\rightarrow(2-\frac{5}{4}Y)^{-1}$ such that at constant mass,
\begin{equation}
    L\propto\frac{1}{(2-\frac{5}{4}Y)^4(2-Y)}.
\end{equation}

\begin{figure}
\centering
\includegraphics[width=0.45\textwidth]{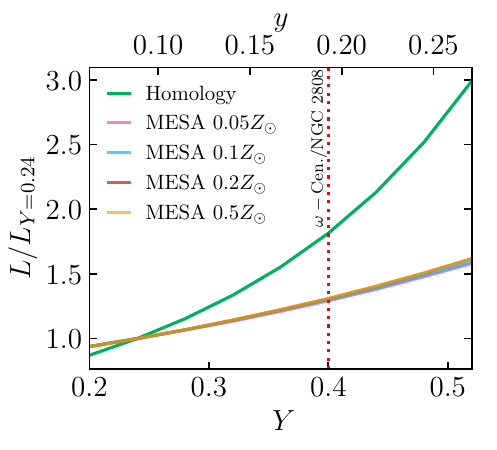}
\caption{(Green) Bolometric luminosity of a fixed-mass star as a function of helium mass fraction ($Y$), scaled to the value assuming a BBN helium abundance ($Y=0.24$). This model assumes that the star is in radiative equilibrium, that the gas is fully-ionized and metal free, and that the opacity is dominated by electron scattering. (Pink, Blue, Brown, Yellow) 1500~\AA\ luminosity integrated over a Salpeter IMF as computed with MESA at $Z=0.05Z_{\odot},\ 0.1Z_{\odot}$, $0.2Z_{\odot}$, and $0.5Z_{\odot}$, respectively, scaled to the value assuming a BBN helium abundance. In both cases, the luminosity is strongly sensitive to $Y$. For reference, the dotted red line indicates $Y=0.4$, which represents the maximum helium abundances seen in Milky Way globular clusters Omega Centauri and NGC~2808.}
\label{fig:l_ratio}
\end{figure}

In Figure~\ref{fig:l_ratio} (green line) we show the bolometric luminosity of a fixed-mass star, subject to the above-listed assumptions, scaled to the luminosity assuming Big Bang Nucleosynthesis (BBN) abundances ($Y=0.24$, \citealt{Planck2020}). For the maximum helium abundance seen in Omega Centauri \citep{Dupree2013} and observed in the $z\sim6$ galaxy RXCJ2248-ID \citep{Topping2024,Yanagisawa2024_He}, the bolometric luminosity increases by a factor of $\sim80\%$.

Stars are of course more complex than this simple model based on homology encapsulates. To gain a more accurate estimate for the luminosity increase specifically at 1500~\AA, we have run a series of {\small MESA} models \citep[][Version r15140]{Paxton2011, Paxton2013, Paxton2015, Paxton2018, Paxton2019} to obtain the effective temperature and bolometric luminosity of the zero-age main-sequence (ZAMS) for different initial metallicities and values of $Y$. Our models supplement those in the literature \citep[e.g.][]{Karakas2014a,Karakas2014b,Chantereau2015,Chantereau2016} that are typically run at a specific metallicity to focus on an individual globular cluster or environment. Models from $1-100\ M_\odot$ were run using the \texttt{pp\_and\_cno\_extras} network and stopped when the hydrogen burning luminosity was within 0.1\% of the total luminosity. Assuming the spectrum of each star follows a blackbody with $T=T_{\rm eff}$, we integrate the total luminosity at $1500$~\AA\ over a \citet{Salpeter1955} IMF with a maximum mass of $100\ {\rm M_{\odot}}$. The pink, blue, and brown, and yellow lines in Figure~\ref{fig:l_ratio} shows the $1500$~\AA\ luminosity as a function of $Y$, normalized by the luminosity at $Y=0.24$ for four different metallicities. Indeed the UV luminosity also increases with $Y$, although less steeply than our simple assumptions predict. Nevertheless, at $Y=0.4$, the UV luminosity is still boosted by 30\%. The relation between helium mass fraction, $Y$, and the luminosity excess at the ZAMS for our assumptions of IMF and $0.1Z_{\odot}$ is well fit by the relation:
\begin{equation}
\label{eqn:lfit}
    L/L_{Y=0.24}=1.94Y^2+0.63Y+0.74.
\end{equation}

Unfortunately, helium abundance measurements are not possible for the highest-redshift nitrogen emitters due to the He$^+$ recombination lines (particularly He\,\textsc{i}\,$\lambda$7065) dropping out of the JWST NIRSpec prism at $z\gtrsim6.5$. To estimate $Y$ for GNz11 \citep{Bunker2023}, GHZ2 \citep{Castellano2024}, and GN-z9p4 \citep{Schaerer2024}, we fit an empirical relation between $\log({\rm N/O})$ and $y={\rm He/H}$ to the data presented in \cite{Yanagisawa2024_He} (their Figure~5) such that $y=0.085e^{(\log({\rm N/O})+1.6)^{2.7}/2.5}$. This gives $y>0.21$ ($Y>0.47$) for GNz11 using the fiducial value of N/O from \cite{Cameron2023_nit}, $0.19<y<0.23$ ($0.43<Y<0.48$) for GHZ2 using the various N/O estimated from \cite{Castellano2024}, and $y=0.13$ ($Y=0.34$) for GN-z9p4 using the N/O measured from \cite{Schaerer2024}. Assuming a Salpeter IMF and a metallicity of $0.1Z_{\odot}$, this would correspond to luminosity boosts of $>43\%$, up to $48\%$, and $18\%$, for GNz11, GHZ2, GN-z9p4, respectively (following Equation~\ref{eqn:lfit}).

\begin{figure}
\centering
\includegraphics[width=0.45\textwidth]{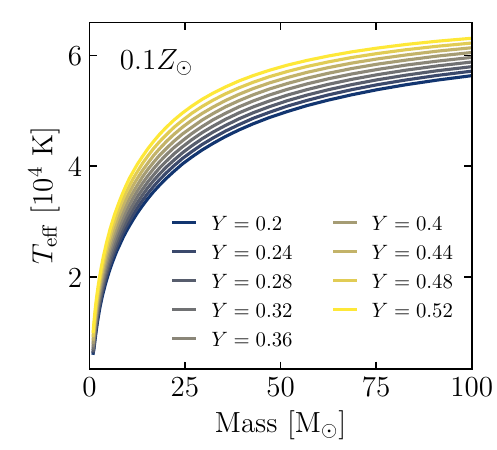}
\caption{Effective temperature as a function of zero-age main-sequence stellar mass for various helium mass fractions at a metallicity of $0.1Z_{\odot}$. Higher values of $Y$ lead to hotter stars. This behavior is consistent across all metallicities in our grid.}
\label{fig:teff}
\end{figure}

Another consequence of an increased helium abundance is that the stellar effective temperature rises. Following from our previous assumptions, homology relations predict $T\propto\mu M/R$, where $R$ is the radius of the star. The increase in temperature is indeed related to the increase in bolometric luminosity. Figure~\ref{fig:teff} highlights the impact of helium enhancement on the ZAMS $T_\mathrm{eff}$ in the same set of {\small MESA} models, which approach $\Delta T_\mathrm{eff}$ of 10~kK for $\gtrsim 25~\mathrm{M_\odot}$ stars for the highest enhancements considered. To first order, hotter, more luminous stars have increased production of ionizing photons and their spectra are harder. It is important to consider that for real stars the exact details of this behavior depend on complex stellar atmosphere physics, which is beyond the scope of this letter. However, a higher ionizing photon production efficiency would lead, in an ionization-bounded nebula, to more nebular continuum emission, which also boosts UV luminosity \citep[e.g.,][]{Raiter2010,Katz2024_BJ}. Moreover, harder photons may help power the observed C~{\small IV} and N~{\small IV}] emission lines, which are not typically seen at such strengths in star-forming galaxies at lower redshifts even at similar metallicities \citep[e.g.,][]{senchynaExtremelyMetalpoorGalaxies2019,bergIntenseIVHe2019,izotovExtremelyStrongIV2024}.

\begin{figure}
\centering
\includegraphics[width=0.45\textwidth]{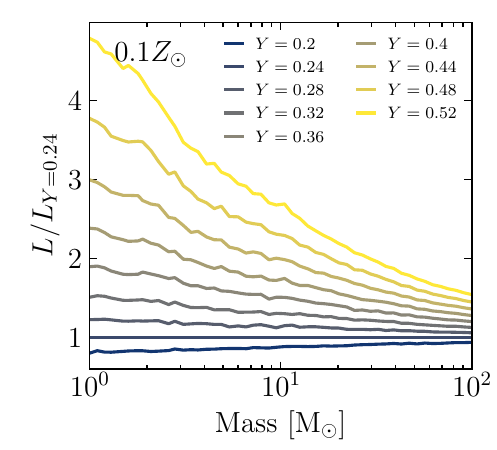}
\caption{Bolometric luminosity of individual stars at the ZAMS as a function of mass for different values of $Y$, normalized to the values at the same mass for $Y=0.24$. Lower-mass stars are more affected by higher values of $Y$. We show results for a metallicity of $0.1Z_{\odot}$; however, other metallicities exhibit similar behavior. Note that the wiggles in the plot are likely numerical due to the finely sampled mass grid.}
\label{fig:lum_by_mass}
\end{figure}

Finally, it is important to highlight that the impact on the total bolometric luminosity is not constant with mass. This is demonstrated in Figure~\ref{fig:lum_by_mass}, where we show the bolometric luminosity of individual stars at the ZAMS as a function of mass for different values of $Y$, normalized to the values at the same mass for $Y=0.24$, assuming a metallicity of $0.1Z_{\odot}$. We observe that lower-mass stars exhibit larger increases in bolometric luminosity for all values of $Y$ compared to higher-mass stars. This effect can be understood from homology relations. The highest mass stars have non-negligible contributions from radiation pressure to their equation of state. For a star entirely dominated by radiation pressure, at fixed mass, $L\propto (2-Y)^{-1}$. At masses $\lesssim2\ {\rm M_{\odot}}$, free-free and bound-free absorption contribute to the opacity where it can be shown that $L\propto \mu^{7.5}$. Because the impact of helium enhancements is mass-dependent, the evolution of properties such as mass-to-light ratio and when spectral features such as the Balmer break appear as a function of time may change and complicate estimates of stellar mass or star formation histories. These again depend on detailed calculations of stellar atmospheres and main-sequence lifetimes, but the effects may be relevant for interpreting the light from early galaxies impacted by globular cluster populations and other densely clustered star formation at a broad range of ages, 
in addition to the very young UV N-emitters that we have focused on in this work.

\section{Discussion \& Conclusions}
\label{sec:conclusions}
Here we have highlighted the fact that helium enhancements, which empirically scale strongly with N/O \citep{Yanagisawa2024_He}, lead to a boost in UV luminosity of individual stars due to the strong scaling of luminosity with mean molecular weight. For the nitrogen emitters observed at $z>10$, detailed stellar evolution models predict that the 1500~\AA\ UV luminosity can increase as much as $\sim50\%$. This effect is likely to take place only in compact star-forming regions undergoing extreme bursts that exhibit similar conditions to what is expected for globular cluster `second population' formation. Such galaxies represent $\sim50\%$ of the very UV-bright population at high redshift, while the others are potentially mergers \citep[e.g.,][]{Harikane2024}. Thus the combination of an intense burst of star formation along with the luminosity increase from an enhanced helium abundance and strong nebular continuum could help explain why there exists a population of compact, very UV-bright galaxies at high redshift. Moreover, if a significant fraction of high-redshift star formation occurs in massive star clusters as predicted from Milky Way data \citep{Belokurov2023}, helium enhancements may be more common at high-redshift than in the local Universe; and could contribute to the general excess UV luminosity density as well as impact retrieval of stellar masses and ages for affected systems.

\vspace{25mm}
\section*{Acknowledgments}
We thank the anonymous referee for their insightful comments that improved the manuscript. HK thanks Richard Ellis, Christopher Lovell, Andrey Kravtsov, and Vasily Belokurov for their comments and discussions on the topic of helium enhancements. We thank the organizers (Volker Bromm, Roberto Maiolino, Brant Robertson, Raffaella Schneider, and Rachel Somerville) of The First Billion Years program hosted at the KITP at the University of California, Santa Barbara, where this research was conceived and conducted. This research was supported in part by grant NSF PHY-2309135 to the Kavli Institute for Theoretical Physics (KITP). 
A.P.J. acknowledges the 2021 MESA Summer School, the University of Chicago's Research Computing Center, and the National Science Foundation grants AST-2206264 and AST-2307599.
OGT acknowledges support from a Carnegie-Princeton Fellowship through Princeton University and the Carnegie Observatories.
P.S.\ acknowledges support from NASA through the NASA Hubble Fellowship grant \#HST-HF2-51565.001-A awarded by the Space Telescope Science Institute, which is operated by the Association of Universities for Research in Astronomy, Incorporated, under NASA contract NAS5-26555.

\bibliographystyle{mn2e}
\bibliography{library_oj}

\end{document}